\documentstyle[12pt]{article}
\newcommand{\Dslash}{\not \!\! D}

\newcommand{\sslash}{\not \!\! s}
\newcommand{\kslash}{\not \! k}
\newcommand{\pslash}{\not \! p}

\begin{document}

\begin{flushright}{UT-02-27}
\end{flushright}
\vskip 0.5 truecm

\begin{center}
{\Large{\bf Supersymmetry on the lattice and the Leibniz 
rule
 }}
\end{center}
\vskip .5 truecm
\centerline{\bf Kazuo Fujikawa}
\vskip .4 truecm
\centerline {\it Department of Physics,University of Tokyo}
\centerline {\it Bunkyo-ku,Tokyo 113,Japan}
\vskip 0.5 truecm

\makeatletter
\@addtoreset{equation}{section}
\def\theequation{\thesection.\arabic{equation}}
\makeatother

\begin{abstract}
The major obstacle to a supersymmetric theory on the 
lattice is the failure of the Leibniz rule. We analyze this 
issue by using the Wess-Zumino model and a general 
Ginsparg-Wilson operator, which is local and free of 
species doublers. We point out that the
Leibniz rule could be maintained on the lattice if the generic 
momentum $k_{\mu}$ carried by any field variable satisfies
$|ak_{\mu}|<\delta$ in the limit $a\rightarrow 0$ for arbitrarily
small but finite $\delta$. This condition is expected to be 
 satisfied generally 
if the theory is finite perturbatively, provided that 
discretization does not induce further symmetry breaking. We 
thus first render the 
continuum Wess-Zumino model finite by applying the higher 
derivative regularization which preserves supersymmetry. We 
then put this theory on the lattice, which preserves 
supersymmetry except for a breaking in interaction terms by 
the failure of the Leibniz rule. By this way, we  
define a lattice Wess-Zumino model which maintains the basic
properties such as $U(1)\times U(1)_{R}$ symmetry and 
holomorphicity. We show that this model reproduces continuum 
theory in the limit $a\rightarrow 0$ up to any finite order 
in perturbation theory; in this sense all the supersymmetry 
breaking terms induced by the failure of the Leibniz rule are 
irrelevant. We then suggest that this discretization 
may work to define a low energy effective theory in a 
non-perturbative way.
\end{abstract}

\section{Introduction}

There are basically two different motivations for defining a 
field
theory on the lattice. The first is to regularize a divergent 
theory and simultaneously define the theory in a 
non-perturbative sense. The second is to define a discretized 
version of a theory, which is finite in continuum perturbation
theory, so that one can apply the numerical and other techniques
 in a 
non-perturbative way. Though this second aspect is not commonly
discussed in the context of lattice theory, we want to 
show that this second aspect may be essential in putting 
 supersymmetric theories\cite{wess-bagger} on the 
lattice\footnote{If one applies 
a discretization to superstring 
theory,for example, it also corresponds to a discretization of 
a perturbatively finite theory.}.

There are several difficulties to define supersymmetry on the 
lattice. The most notable and difficult issue is the failure 
of the Leibniz rule\cite{dondi}. To be explicit, we have on the 
lattice
\begin{eqnarray}
&&\frac{1}{a}(f(x+a)g(x+a)-f(x)g(x))
\nonumber\\
&&=\frac{1}{a}(f(x+a)-f(x))g(x)+f(x)\frac{1}{a}(g(x+a)-g(x))
\nonumber\\
&&+a\frac{1}{a}(f(x+a)-f(x))\frac{1}{a}(g(x+a)-g(x))
\end{eqnarray}
namely the ``lattice version of the Leibniz rule'' is given 
by\footnote{If one uses the symmetric difference, 
\begin{eqnarray}
&&(\nabla (fg))(x)=(f(x+a)g(x+a)-f(x-a)g(x-a))/(2a)\nonumber\\
&&=(\nabla f)(x)g(x)+f(x)(\nabla g)(x)+a(\nabla f)(x)
[(g(x+a)-g(x))/a]
+a[(f(x)-f(x-a))/a](\nabla g)(x)\nonumber
\end{eqnarray}
and still the limit $a\rightarrow 0$ is not smoothly defined
in general. }
\begin{equation}
(\nabla (fg))(x)=(\nabla f)(x)g(x)+f(x)(\nabla g)(x)
+a(\nabla f)(x)(\nabla g)(x).
\end{equation}
This shows that the breaking of supersymmetry by the lattice 
artifact is formally of order $O(a)$, but actually the 
breaking is of order $O(1)$ if the momentum carried by the 
field variables is of order $O(1/a)$. To recover the 
conventional Leibniz rule,  a necessary  condition for the 
momentum variable is
\begin{equation}
|ak_{\mu}| <\delta
\end{equation}
for $a\rightarrow 0$ with arbitrarily small but finite 
$\delta$. Here $k_{\mu}$ is a generic momentum carried 
by any field variable in the Feynman diagrams so that the last 
term in the lattice Leibniz rule (1.2) is neglected to give
\begin{equation}
(\nabla (fg))(x)=(\nabla f)(x)g(x)+f(x)(\nabla g)(x).
\end{equation}
This requirement is expected to be satisfied if the theory in 
continuum is 
finite  in a perturbative sense so that all the momentum 
variables in Feynman diagrams are finite and thus 
infinitesimally small measured by the lattice unit $1/a$ in the 
limit $a\rightarrow 0$, provided that the lattice 
discretization 
does not introduce further symmetry breaking terms. 

If the above condition is satisfied, all the supersymmetry 
breaking 
terms for finite lattice spacing, which are induced by the 
failure of the Leibniz rule,  are expected to be irrelevant 
in the sense 
that those supersymmetry breaking terms vanish in the limit 
$a\rightarrow 0$. In the context of supersymmetric Yang-Mills
theory on the lattice, an argument to the effect that 
all the supersymmetry 
breaking terms are irrelevant was given in the past\cite{curci}
, though
the $N=1$ supersymmetric Yang-Mills theory is not finite
and thus the basic reasoning is completely different.
In the context of the Wess-Zumino model\cite{wess}, 
we would like to show that a
sensible lattice discretization, which is based on the presently
available technique, may be to first render the continuum 
Wess-Zumino model finite by applying the higher derivative 
regularization. This higher derivative regularization is known 
to preserve supersymmetry in a perturbative 
sense\cite{iliopoulos}. We then apply the lattice 
discretization to this regularized continuum theory. 

In this paper, we present a detailed analysis of the above 
 procedure 
for the Wess-Zumino model in the framework of perturbation 
theory, and then suggest that this scheme may work in a 
non-perturbatice sense also. 
We utilize the Ginsparg-Wilson fermion operators which are 
local and free of species 
doublers\cite{ginsparg}-\cite{niedermayer}. In the course of 
this analysis,  we clarify some of the subtleties appearing 
in the Ginsparg-Wilson operators when utilized in the present 
context.

\section{Wess-Zumino model on the lattice}

The Wess-Zumino model is the simplest supersymmetric model
in 4-dimensional space time, and  the non-renormalization 
theorem was first discovered in this model: If renormalized
at vanishing momenta, all the potential terms including mass
terms do not receive any (even finite) renormalization, except 
for a uniform wave function renormalization, up to 
all orders in perturbation theory\cite{fujikawa2}.

As for the previous studies of the Wess-Zumino model on the 
lattice, see, for example, \cite{bartels}-\cite{kikukawa}.

\subsection{Lattice Lagrangian}

We define the Wess-Zumino model on the lattice 
by\cite{fuji-ishi1}
\begin{eqnarray}
{\cal L}&=&\frac{1}{2}\chi^{T}C\frac{1}{\Gamma_{5}}
\gamma_{5}D\chi + \frac{1}{2}m\chi^{T}C\chi
+ g\chi^{T}C(P_{+}\phi P_{+}+P_{-}\phi^{\dagger}P_{-})\chi
\nonumber\\
&&-\phi^{\dagger}D^{\dagger}D\phi
+F^{\dagger}\frac{1}{\Gamma^{2}_{5}}F
+m[F\phi+(F\phi)^{\dagger}]
+g[F\phi^{2}+(F\phi^{2})^{\dagger}]\nonumber\\
&=&\frac{1}{2}\chi^{T}C\frac{1}{\Gamma_{5}}
\frac{H}{a}\chi + \frac{1}{2}m\chi^{T}C\chi
+ g\chi^{T}C(P_{+}\phi P_{+}+P_{-}\phi^{\dagger}P_{-})\chi
\nonumber\\
&&-\phi^{\dagger}\frac{H^{2}}{a^{2}}\phi
+F^{\dagger}\frac{1}{\Gamma^{2}_{5}}F
+m[F\phi+(F\phi)^{\dagger}]
+g[F\phi^{2}+(F\phi^{2})^{\dagger}].
\end{eqnarray}
Here the hermitian operator 
\begin{equation}
H=a\gamma_{5}D=H^{\dagger}
\end{equation}
satisfies the general Ginsparg-Wilson relation\cite{fujikawa}
\begin{equation}
\gamma_{5}H+H\gamma_{5}=2H^{2k+2}
\end{equation}
with a non-negative integer $k$, which implies 
\begin{equation}
\gamma_{5}H^{2}=(\gamma_{5}H+H\gamma_{5})H
-H(\gamma_{5}H+H\gamma_{5})+H^{2}\gamma_{5}=H^{2}\gamma_{5},
\end{equation}
and 
\begin{equation}
\Gamma_{5}=\gamma_{5}-H^{2k+1}
\end{equation}
which satisfies $H\Gamma_{5}+\Gamma_{5}H=0$. 
An explicit form of $H$ is given in Appendix.
When we have $H^{2}$ and $\Gamma^{2}_{5}$ in 
the bosonic terms, we adopt the convention to discard
 the unit Dirac matrix in $H^{2}$ (see Appendix). Note that 
$\Gamma_{5}^{2}=1-H^{4k+2}$, and $H^{2}=1$ just on top of the 
would-be species doublers. 

The projection operators are defined by\cite{niedermayer} 
\begin{equation}
P_{\pm}=\frac{1}{2}(1\pm\gamma_{5}),\ \ \ \ \ 
\hat{P}_{\pm}=\frac{1}{2}(1\pm \hat{\gamma}_{5})
\end{equation}
with $\hat{\gamma}_{5}=\gamma_{5}-2H^{2k+1}$ which satisfies
$\hat{\gamma}^{2}_{5}=1$.

A salient feature of our Lagrangian is that it is invariant 
under the continuum chiral transformation, except for the 
mass term, if one performs simultaneously a suitable phase 
rotation of the 
fields $\phi$ and $\phi^{\dagger}$. 
We here note the relation which 
follows from the defining relation of $H$ (see also 
Appendix)
\begin{equation}
\Gamma_{5}H/a=\frac{1}{a}(\gamma_{5}H-H^{2+2k})
=\frac{1}{2a}[\gamma_{5},H]
\propto \gamma^{\mu}\frac{\sin ap_{\mu}}{a}.
\end{equation}
We then have 
\begin{equation}
\{\gamma_{5},\Gamma_{5}H\}
=\frac{1}{2}\{\gamma_{5},[\gamma_{5},H]\}=0
\end{equation} 
which suggests that the fermion kinetic operator satisfies
\begin{equation}
\{\gamma_{5}, \frac{1}{\Gamma_{5}}H\}
=\{\gamma_{5}, \frac{1}{\Gamma^{2}_{5}}\Gamma_{5}H \}=0
\end{equation}
by using $[\gamma_{5},\Gamma^{2}_{5}]=0$.
The factor $\Gamma_{5}$ in the fermion kinetic 
term\footnote{This structure of the fermion kinetic term is 
required to define the Euclidean Majorana fermion\cite{nicolai} 
in a consistent manner in the presence of chiral symmetric 
Yukawa couplings\cite{fuji-ishi1}.} $(1/\Gamma_{5})H$ 
vanishes at the momentum 
corresponding to the would-be species 
doublers \cite{fuji-ishi1}\cite{fis}, but this non-locality
is compensated for in the present supersymmetric theory by the 
corresponding singularity in the term $F^{\dagger}F$. 
Note that variables $\{\chi,\phi,F\}$ are treated as components
of a single superfield in supersymmetric theory. In fact 
one can confirm that the partition function of the free part 
of the Lagrangian gives  unity and thus the factor $\Gamma_{5}$
is cancelled among the component fields:
\begin{eqnarray}
&&\int{\cal D}\chi{\cal D}\phi{\cal D}\phi^{\dagger}
{\cal D}F{\cal D}F^{\dagger}\exp\{\int[
\frac{1}{2}\chi^{T}C\frac{1}{\Gamma_{5}}\frac{H}{a}\chi 
+ \frac{1}{2}m\chi^{T}C\chi
\nonumber\\
&&\qquad-\phi^{\dagger}\frac{H^{2}}{a^{2}}\phi
+F^{\dagger}\frac{1}{\Gamma^{2}_{5}}F
+m[F\phi+(F\phi)^{\dagger}]]\}\nonumber\\
&&=\int{\cal D}\chi{\cal D}\phi{\cal D}\phi^{\dagger}
{\cal D}F^{\prime}{\cal D}(F^{\prime})^{\dagger}\exp\{\int[
\frac{1}{2}\chi^{T}C\frac{1}{\Gamma_{5}}\frac{H}{a}\chi 
+ \frac{1}{2}m\chi^{T}C\chi
\nonumber\\
&&\qquad-\phi^{\dagger}\frac{H^{2}}{a^{2}}\phi
-\phi^{\dagger}(m\Gamma_{5})^{2}\phi
+(F^{\prime})^{\dagger}\frac{1}{\Gamma^{2}_{5}}F^{\prime}]\}
\nonumber\\
&&=\frac{\sqrt{\det[\frac{1}{\Gamma_{5}}\frac{H}{a}+m]}}
{\det[\frac{H^{2}}{a^{2}}+(m\Gamma_{5})^{2}]
\det[\frac{1}{\Gamma^{2}_{5}}]}
=\frac{\sqrt{\det[\frac{1}{\Gamma_{5}}]
\det[\frac{H}{a}+m\Gamma_{5}]}}
{\det[\frac{H^{2}}{a^{2}}+(m\Gamma_{5})^{2}]
\det[\frac{1}{\Gamma^{2}_{5}}]}\nonumber\\
&&=\frac{\{\det[\frac{1}{\Gamma^{2}_{5}}]
\det[(\frac{H}{a})^{2}+(m\Gamma_{5})^{2}]\}^{1/4}}
{\det[\frac{H^{2}}{a^{2}}+(m\Gamma_{5})^{2}]
\det[\frac{1}{\Gamma^{2}_{5}}]}=1
\end{eqnarray}
if one recalls that both of  $\Gamma^{2}_{5}$ and 
$(\frac{H}{a})^{2}$ are proportional to a $4\times 4$ unit 
matrix and that this unit matrix is neglected in the bosonic 
sector appearing in the denominator.

We shall also confirm 
that perturbation theory is well defined without any 
singularity, though  the fermion propagator vanishes at the 
momenta corresponding to the would-be species doublers. In the 
non-perturbative formulation, the factor $1/\Gamma_{5}$
may be compensated for  by rescaling the variables $F$ and 
$F^{\dagger}$, as will be shown later.
 
Our convention of the charge conjugation matrix is 
\begin{eqnarray}
&&C\gamma^{\mu}C^{-1}=-(\gamma^{\mu})^{T},\\ 
&&C\gamma_{5}C^{-1}=\gamma^{T}_{5},\\ 
&&C^{\dagger}C=1,\ \  C^{T}=-C
\end{eqnarray}
and the Ginsparg-Wilson operator satisfies
\begin{eqnarray}
&&C\gamma_{5}\Gamma_{5}C^{-1}=
(\gamma_{5}\Gamma_{5})^{T},
\nonumber\\
&&CDC^{-1}=D^{T}.
\end{eqnarray}

\subsection{Supersymmetry}

If one defines the real components by 
\begin{equation}
\phi\rightarrow\frac{1}{\sqrt{2}}(A+iB), \ \ \ F\rightarrow 
\frac{1}{\sqrt{2}}(F-iG)
\end{equation}
the above Lagrangian is written as 
\begin{eqnarray}
{\cal L}
&=&\frac{1}{2}\chi^{T}C\frac{1}{\Gamma_{5}}
\frac{1}{a}H\chi
-\frac{1}{2a^{2}}[AH^{2}A+BH^{2}B]\nonumber\\
&&+\frac{1}{2}[F\frac{1}{\Gamma^{2}_{5}}F
+G\frac{1}{\Gamma^{2}_{5}}G]\nonumber\\
&&+\frac{1}{2}m\chi^{T}C\chi
+m[FA+GB]\nonumber\\
&&\frac{1}{\sqrt{2}}g\chi^{T}C(A+i\gamma_{5}B)\chi
+\frac{1}{\sqrt{2}}g[F(A^{2}-B^{2})+2G(AB)].
\end{eqnarray}
The free part of the action formed from this Lagrangian is 
confirmed to be invariant 
under the ``lattice supersymmetry'' transformation\footnote{
We assume that our kinetic operator $\Gamma_{5}H$ satisfies the 
proper charge conjugation symmetry $C\Gamma_{5}HC^{-1}
=(\Gamma_{5}H)^{T}$, which includes an operation 
corresponding to partial integration.}
\begin{eqnarray}
&&\delta\chi=-\Gamma_{5}\frac{1}{a}H
(A-i\gamma_{5}B)\epsilon-(F-i\gamma_{5}G)\epsilon ,
\nonumber\\
&&\delta A=\epsilon^{T}C\chi=\chi^{T}C\epsilon,\nonumber\\
&&\delta B=-i\epsilon^{T}C\gamma_{5}\chi=-i\chi^{T}C\gamma_{5}
\epsilon\nonumber\\
&&\delta F=\epsilon^{T}C\Gamma_{5}\frac{1}{a}H\chi
\sim\chi^{T}C\Gamma_{5}\frac{1}{a}H\epsilon\nonumber\\
&&\delta G=i\epsilon^{T}C\Gamma_{5}\frac{1}{a}H\gamma_{5}\chi
\sim -i\chi^{T}C\Gamma_{5}\frac{1}{a}H\gamma_{5}\epsilon
\end{eqnarray}
with a constant Majorana-type Grassmann parameter $\epsilon$.
Note that the order of the operators is important in these
expressions. The second expressions of $\delta F$ and 
$\delta G$ need to be treated carefully when these variations 
are multiplied with other field variables\footnote{For example,
in the presence of a scalar field $A(x)$, we have 
$\int A\epsilon^{T}C\Gamma_{5}\frac{1}{a}H\chi=\int
\chi^{T}C\Gamma_{5}\frac{1}{a}H(A\epsilon)$. }.
If one recalls the correspondence to continuum theory in the 
naive limit $a\rightarrow 0$
\begin{eqnarray}
&&\frac{1}{a}H\leftrightarrow \gamma_{5}\Dslash\nonumber\\
&&\Gamma_{5}\leftrightarrow \gamma_{5},
\end{eqnarray} 
the above transformation defines a lattice generalization of the 
continuum supersymmetry 
transformation\cite{wess}\cite{iliopoulos}.

For example, the variation of the kinetic terms under the above 
transformation is given by
\begin{eqnarray}
&&\delta\int{\cal L}_{kin}\nonumber\\
&=&\int\{\chi^{T}C\frac{1}{\Gamma_{5}}
\frac{1}{a}H[-\Gamma_{5}\frac{1}{a}H
(A-i\gamma_{5}B)\epsilon-(F-i\gamma_{5}G)\epsilon]\nonumber\\
&&-\frac{1}{a^{2}}AH^{2}[\epsilon^{T}C\chi]
-\frac{1}{a^{2}}BH^{2}[-i\epsilon^{T}C\gamma_{5}\chi]
\nonumber\\
&&+F\frac{1}{\Gamma^{2}_{5}}
[\epsilon^{T}C\Gamma_{5}\frac{1}{a}H\chi]
+G\frac{1}{\Gamma^{2}_{5}}[
i\epsilon^{T}C\Gamma_{5}\frac{1}{a}H\gamma_{5}\chi]\}\nonumber\\
&=&\int\{\chi^{T}C
\frac{1}{a^{2}}H^{2}(A-i\gamma_{5}B)\epsilon
-\chi^{T}C\frac{1}{\Gamma_{5}}
\frac{1}{a}H(F-i\gamma_{5}G)\epsilon\nonumber\\
&&-\frac{1}{a^{2}}\chi^{T}C H^{2}(A-i\gamma_{5}B)\epsilon
+\chi^{T}C\frac{1}{\Gamma^{2}_{5}}\Gamma_{5}
\frac{1}{a}H(F-i\gamma_{5}G)\epsilon\}
=0
\end{eqnarray}
which is consistent with (2.10). Here we used (2.8).

The lattice supersymmetry variation of the interaction terms is 
given by
\begin{eqnarray}
&&\delta\int{\cal L}_{int}\nonumber\\
&=&\int\{\frac{2}{\sqrt{2}}g\chi^{T}C
(A+i\gamma_{5}B)[-\Gamma_{5}\frac{1}{a}H
(A-i\gamma_{5}B)\epsilon-(F-i\gamma_{5}G)\epsilon]\nonumber\\
&&+\frac{2}{\sqrt{2}}g\{FA[\epsilon^{T}C\chi]
-FB[-i\epsilon^{T}C\gamma_{5}\chi]
+2GA[-i\epsilon^{T}C\gamma_{5}\chi]
+2GB[\epsilon^{T}C\chi]\}
\nonumber\\
&&+\frac{1}{\sqrt{2}}g
\{(A^{2}-B^{2})[\epsilon^{T}C\Gamma_{5}\frac{1}{a}H\chi]
+2(AB)[i\epsilon^{T}C\Gamma_{5}\frac{1}{a}H\gamma_{5}\chi]\}\}
\nonumber\\
&=&\int\{\frac{2}{\sqrt{2}}g\chi^{T}C
(A+i\gamma_{5}B)[-\Gamma_{5}\frac{1}{a}H
(A-i\gamma_{5}B)\epsilon]\nonumber\\
&&+\frac{1}{\sqrt{2}}g
\{(A^{2}-B^{2})[\epsilon^{T}C\Gamma_{5}\frac{1}{a}H\chi]
+2(AB)[i\epsilon^{T}C\Gamma_{5}\frac{1}{a}H\gamma_{5}\chi]\}\}
\nonumber\\
&=&\int\{\frac{2}{\sqrt{2}}g\chi^{T}C
(A+i\gamma_{5}B)[-\Gamma_{5}\frac{1}{a}H
(A-i\gamma_{5}B)\epsilon]\nonumber\\
&&+\frac{1}{\sqrt{2}}g
\{[\chi^{T}C\Gamma_{5}\frac{1}{a}H(A^{2}-B^{2})\epsilon]
-2[i\chi^{T}C\Gamma_{5}\frac{1}{a}H\gamma_{5}(AB)\epsilon]\}\}
\nonumber\\
&=&-\int\{\frac{2}{\sqrt{2}}g\chi^{T}C
(A\Gamma_{5}\frac{1}{a}HA-B\Gamma_{5}\frac{1}{a}HB)\epsilon
\nonumber\\
&&+\frac{2i}{\sqrt{2}}g\chi^{T}C[A\Gamma_{5}\frac{1}{a}HB
+B\Gamma_{5}\frac{1}{a}HA]\gamma_{5}\epsilon\nonumber\\
&&+\frac{1}{\sqrt{2}}g
\{[\chi^{T}C\Gamma_{5}\frac{1}{a}H(A^{2}-B^{2})\epsilon]
-2i[\chi^{T}C\Gamma_{5}\frac{1}{a}H(AB)\gamma_{5}\epsilon]\}\}.
\end{eqnarray}
Here we used the relation
$\gamma_{5}\Gamma_{5}H=-\Gamma_{5}H\gamma_{5}$ (2.8).
If the operator $\Gamma_{5}H/a$ (2.7) satisfies the Leibniz 
rule, the above  variation of the interaction terms vanishes.
We thus  encounter the notorious issue related to the Leibniz 
rule, which is basically the lattice artifact.
 
The propagators for perturbative calculations are given by
\begin{eqnarray}
&&\langle\phi\phi^{\dagger}\rangle=\frac{a^{2}}
{H^{2}+(am\Gamma_{5})^{2}}\nonumber\\
&&\langle F F^{\dagger}\rangle=(-)\frac{H^{2}\Gamma^{2}_{5}}
{H^{2}+(am\Gamma_{5})^{2}}\nonumber\\
&&\langle F \phi\rangle
=\langle F^{\dagger} \phi^{\dagger}\rangle=(-)
\frac{a^{2}m\Gamma^{2}_{5}}
{H^{2}+(am\Gamma_{5})^{2}}\nonumber\\
&&\langle\chi(y)\chi^{T}(x)C\rangle=(-)\frac{a}{H+am\Gamma_{5}}
\Gamma_{5}=(-)\gamma_{5}\Gamma_{5}\frac{a}{H+am\Gamma_{5}}
\gamma_{5}
\end{eqnarray}
and other propagators vanish. When we have $H^{2}$ and 
$\Gamma^{2}_{5}$ in 
the bosonic propagators, we adopt the convention to discard
 the unit Dirac matrix in $H^{2}$. Note that 
$\Gamma_{5}^{2}=1-H^{4k+2}$.
If one uses the identities 
\begin{eqnarray}
&&P_{+}\phi(x)\hat{P}_{+}=P_{+}\phi(x)P_{+}
\gamma_{5}\Gamma_{5},\nonumber\\
&&P_{-}\phi^{\dagger}(x)\hat{P}_{-}=P_{-}\phi^{\dagger}(x)
P_{-}\gamma_{5}\Gamma_{5},
\end{eqnarray}
the Feynman rules in the present scheme are essentially 
identical to those in the 
previous calculation\cite{fuji-ishi2}. The one-loop level 
non-renormalization theorem when remormalized at vanishing 
momenta is thus satisfied, though the kinetic 
terms receive non-uniform finite renormalization in addtion to 
 uniform logarithmic renormalization\cite{fuji-ishi2}. 

The holomorphic properties in a naive sense are preserved in 
our Lagrangian\footnote{The decisive factor to ensure the
non-renormalization theorem is supersymmetry, while 
$U(1)\times U_{R}(1)$ symmetry and holomorphicity provide 
additional constraints.}.  
As for the 
$U(1)\times U_{R}(1)$ charges, 
where $U_{R}(1)$ stands for the R-symmetry, we first write the 
potential part of the Lagrangian as
\begin{eqnarray}
{\cal L}_{pot}&=&\frac{1}{2}m(P_{+}\chi)^{T}CP_{+}\chi+mF\phi
+ \chi^{T}CP_{+}g\phi P_{+}\chi+gF\phi^{2}\\
&&+\frac{1}{2}m^{\dagger}(P_{-}\chi)^{T}CP_{-}\chi
+(mF\phi)^{\dagger}
+ \chi^{T}CP_{-}(g\phi)^{\dagger} P_{-}\chi
+(gF\phi^{2})^{\dagger}\nonumber
\end{eqnarray}
and we assign\cite{seiberg}
\begin{eqnarray}
&&\phi = (1,1),\nonumber\\
&&F = (1 ,-1),\nonumber\\
&&P_{+}\chi = \xi = (1,0),\nonumber\\
&&m = (-2,0),\nonumber\\
&&g = (-3,-1)
\end{eqnarray}
by regarding $m$ and $g$ as complex parameters. Here 
$\xi$ is a two component spinor in the representation 
where $\gamma_{5}$ is diagonal.    
For $m=g=0$, our Lagrangian preserves these charges if one 
recalls 
\begin{equation}
\frac{1}{\Gamma_{5}}H
=P_{-}\frac{1}{\Gamma_{5}}H P_{+}+
P_{+}\frac{1}{\Gamma_{5}}H P_{-}
\end{equation}
and  $P_{-}\chi=\xi^{\star}$ in the representation where 
$\gamma_{5}$ is diagonal.

\section{Higher derivative regularization on the lattice}

In the perturbative treatment of the above lattice Lagrangian 
(2.1), 
the higher order diagrams generally break supersymmetry 
because all the momentum regions contribute to the loop 
diagrams; it is not easy to preserve supersymmetry in this 
strict sense ( i.e., for all the momentum regions ) on the 
lattice because of the failure of the Leibniz rule. 

A way to resolve this difficulty may be to apply a higher 
derivative regularization on the lattice. By this way, one can
transfer all the divergences to the infrared divergences 
measured by the lattice unit $1/a$.
In the infrared region, the momenta in loop 
diagrams are constrained to the momentum region in continuum 
theory, for which the lattice artifact such as the failure of
the Leibniz rule  could be negligible in the limit 
$a\rightarrow 0$.

The higher derivative regularization in the present lattice 
Lagrangian is implemented by 
\begin{eqnarray}
{\cal L}&=&\frac{1}{2}\chi^{T}C\frac{1}{\Gamma_{5}}
\gamma_{5}D\frac{(D^{\dagger}D+M^{2})}{M^{2}}\chi 
+ \frac{1}{2}m\chi^{T}C\frac{(D^{\dagger}D+M^{2})}{M^{2}}
\chi\nonumber\\
\nonumber\\
&&-\phi^{\dagger}D^{\dagger}D\frac{(D^{\dagger}D+M^{2})}{M^{2}}
\phi
+F^{\dagger}\frac{1}{\Gamma^{2}_{5}}
\frac{(D^{\dagger}D+M^{2})}{M^{2}}F
\nonumber\\
&&+m[F\frac{(D^{\dagger}D+M^{2})}{M^{2}}\phi
+(F\frac{(D^{\dagger}D+M^{2})}{M^{2}}\phi)^{\dagger}]
\nonumber\\
&&+ g\chi^{T}C(P_{+}\phi P_{+}+P_{-}\phi^{\dagger}P_{-})\chi
+g[F\phi^{2}+(F\phi^{2})^{\dagger}]\nonumber\\
&=&\frac{1}{2}\chi^{T}C\frac{1}{\Gamma_{5}}
\frac{H}{a}\frac{(H^{2}+(aM)^{2})}{(aM)^{2}}\chi 
+ \frac{1}{2}m\chi^{T}C\frac{(H^{2}+(aM)^{2})}{(aM)^{2}}
\chi
\nonumber\\
&&-\phi^{\dagger}\frac{H^{2}}{a^{2}}
\frac{(H^{2}+(aM)^{2})}{(aM)^{2}}\phi
+F^{\dagger}\frac{1}{\Gamma^{2}_{5}}
\frac{(H^{2}+(aM)^{2})}{(aM)^{2}}F
\\
&&+m[F\frac{(H^{2}+(aM)^{2})}{(aM)^{2}}\phi
+(F\frac{(H^{2}+(aM)^{2})}{(aM)^{2}}\phi)^{\dagger}]
\nonumber\\
&&+ g\chi^{T}C(P_{+}\phi P_{+}+P_{-}\phi^{\dagger}P_{-})\chi
+g[F\phi^{2}+(F\phi^{2})^{\dagger}].\nonumber
\end{eqnarray}
The $U(1)\times U_{R}(1)$ symmetry and holomorphicity are 
preserved in this regularization.

The propagators for perturbative calculations are given by
\begin{eqnarray}
&&\langle\phi\phi^{\dagger}\rangle=\frac{a^{2}}
{H^{2}+(am\Gamma_{5})^{2}}\frac{(aM)^{2}}{H^{2}+(aM)^{2}}
\nonumber\\
&&\langle F F^{\dagger}\rangle=(-)\frac{H^{2}\Gamma^{2}_{5}}
{H^{2}+(am\Gamma_{5})^{2}}\frac{(aM)^{2}}{H^{2}+(aM)^{2}}
\nonumber\\
&&\langle F \phi\rangle
=\langle F^{\dagger} \phi^{\dagger}\rangle=(-)
\frac{a^{2}m\Gamma^{2}_{5}}
{H^{2}+(am\Gamma_{5})^{2}}\frac{(aM)^{2}}{H^{2}+(aM)^{2}}
\nonumber\\
&&\langle\chi(y)\chi^{T}(x)C\rangle=(-)\gamma_{5}\Gamma_{5}
\frac{a}{H+am\Gamma_{5}}
\frac{(aM)^{2}}{H^{2}+(aM)^{2}}\gamma_{5}\nonumber\\
&&=
\frac{a\Gamma_{5}H-a^{2}m(\Gamma_{5})^{2}}
{H^{2}+(am\Gamma_{5})^{2}}
\frac{(aM)^{2}}{H^{2}+(aM)^{2}}
\end{eqnarray}
and other propagators vanish. When we have $H^{2}$ and 
$\Gamma^{2}_{5}$ in 
the bosonic sector, we adopt the convention to discard
 the unit Dirac matrix in $H^{2}$. Note that 
$\Gamma_{5}H+H\Gamma_{5}=0$ and 
$\Gamma_{5}^{2}=1-H^{4k+2}$.
Here $M$ is a new mass scale which may be chosen to be
\begin{equation}
1\gg (aM)^{2}\gg (am)^{2}.
\end{equation}

One can confirm that the free part of this Lagrangian with
higher derivative regularization (3.1) is still invariant under 
the lattice supersymmetry transformation (2.17) by noting 
$[\Gamma_{5},H^{2}]=0$, while the interaction 
terms are not modified by the higher derivative regularization.
 
\subsection{One-loop tadpole and self-energy corrections}

It has been shown previously that the superpotential is not 
renormalized in the one-loop level even for a finite $a$ when 
renormalized at vanishing momenta\cite{fuji-ishi2}. This 
conclusion still holds in the present 
model with higher derivative regularization. 
One can also confirm that the cancellation of tadpole
diagrams is still maintained even for a finite $a$ 
in the one-loop level in the present model\footnote{Intuitively,
this one-loop cancellation arises from the fact that the 
interaction terms, when one of $\phi$ or $\phi^{\dagger}$ is 
set to a constant, are reduced to the effective mass terms which 
are invariant under lattice supersymmetry. }.

It is instructive to analyze the tadpole diagrams in some 
detail. The scalar tadpole contribution to the fields $\phi$ and 
$\phi^{\dagger}$ is given
by 
\begin{eqnarray}
&&2g[\langle F\phi\rangle\phi+\langle F^{\dagger}
\phi^{\dagger}\rangle\phi^{\dagger}]\nonumber\\
&&=-2g\int[\phi\frac{a^{2}m\Gamma_{5}^{2}}
{H^{2}+(am\Gamma_{5})^{2}}\frac{(aM)^{2}}{H^{2}+(aM)^{2}}
+\phi^{\dagger}\frac{a^{2}m\Gamma^{2}_{5}}
{H^{2}+(am\Gamma_{5})^{2}_{5}}\frac{(aM)^{2}}{H^{2}+(aM)^{2}}]
\nonumber\\
&&=-2mga^{2}(\phi+\phi^{\dagger})\int\frac{\Gamma_{5}^{2}}
{H^{2}+(am\Gamma_{5})^{2}}\frac{(aM)^{2}}{H^{2}+(aM)^{2}}\nonumber\\
&&=-2mg(\phi+\phi^{\dagger})\int_{-\pi}^{\pi}
\frac{d^{4}k}{(2\pi)^{4}}
\frac{\Gamma_{5}^{2}(k)}
{H^{2}(k)+(am\Gamma_{5}(k))^{2}}
\frac{M^{2}}{H^{2}(k)+(aM)^{2}}
\end{eqnarray} 
where we chose the basic Brillouin zone at 
\begin{equation}
\frac{-\pi}{a}< k_{\mu}\leq \frac{\pi}{a}
\end{equation}
and rescaled the integration variable
as $a k_{\mu}\rightarrow k_{\mu}$. If one considers the limit 
$a\rightarrow 0$ in the above integral, one obtains a logarithmic
divergence from the region $k_{\mu}\sim 0$. But it is important 
to recognize that the entire region of the basic Brillouin zone
gives a finite contribution in the above integral even at the 
limit $a\rightarrow 0$.
This is the peculiar feature of the one-loop tadpole diagrams 
in the present minimal higher derivative 
regularization\footnote{If one considers the higher derivative 
regularization with the factor 
$((H^{2}+(aM)^{2})/(aM)^{2})^{2}$ instead of 
$(H^{2}+(aM)^{2})/(aM)^{2}$ in (3.1), even the one-loop 
tadpole diagrams receive non-vanishing contributions only 
from the infrared region in the limit $a\rightarrow 0$. This 
stronger regularization may be necessary for a numerical 
simulation.}, and 
all other diagrams receive non-vanishing contributions only 
from the infrared region $k_{\mu}\sim 0$ in the limit 
$a\rightarrow 0$. In any case, it is confirmed that the 
above scalar tadpole contribution is precisely cancelled by a 
fermion tadpole contribution even for a finite $a$.

Since the renormalization of kinetic terms was not uniform 
{\em without} the higher derivative 
regularization\cite{fuji-ishi2}, 
we here analyze the kinetic terms and associated quadratic and 
logarithmic divergences in the one-loop level in more detail. 
The one-loop correction to the ``kinetic term'' $FF^{\dagger}$
is given by
\begin{eqnarray}
&&\frac{g^{2}}{2!}[F(\phi)^{2}
+(F(\phi)^{2})^{\dagger}]^{2}\nonumber\\
&&\rightarrow
\frac{g^{2}}{2!}[4F\langle\phi\phi^{\dagger}\rangle
\langle\phi\phi^{\dagger}\rangle F^{\dagger}]
\nonumber\\
&&=2g^{2}a^{4}F[\int
\frac{1}{H^{2}+(am\Gamma_{5})^{2}}\frac{(aM)^{2}}{H^{2}+(aM)^{2}}
\nonumber\\
&&\times\frac{1}{H^{2}+(am\Gamma_{5})^{2}}
\frac{(aM)^{2}}{H^{2}+(aM)^{2}}]F^{\dagger}
\end{eqnarray}
which is logarithmically divergent.

The integral for $FF^{\dagger}$  is written in more 
detail as
\begin{eqnarray}
&&2g^{2}[\int_{-\pi}^{\pi}\frac{d^{4}k}{(2\pi)^{4}} 
\frac{1}{H^{2}(k+ap)+(am\Gamma_{5})^{2}(k+ap)}
\frac{(aM)^{2}}{H^{2}(k+ap)+(aM)^{2}}\nonumber\\
&&\times\frac{1}{H^{2}(k)+(am\Gamma_{5})^{2}(k)}
\frac{(aM)^{2}}{H^{2}(k)+(aM)^{2}}]
\end{eqnarray}
where we rescaled the integration variable
as $a k_{\mu}\rightarrow k_{\mu}$ by choosing the 
basic Brillouin zone as in (3.5). 
In this integral, if one 
chooses the integration domain {\em outside} the infrared 
region
\begin{equation}
\delta >k_{\mu}>-\delta \ \ \ \ for\ \ \  all \ \ \ \mu
\end{equation}
for arbitrarily small but finite $\delta$,
the integral vanishes for $a\rightarrow 0$ since we have no
infrared divergences\footnote{This and following analyses are 
extended to $\delta=(aM)^{\epsilon}$ with  sufficiently 
small positive $\epsilon$. }. In this analysis, the absence of 
species doubling in the Ginsparg-Wilson operator is 
essential: Namely, $H^{2}\sim 1$ for the momentum domain 
of the would-be species doublers. 

The non-vanishing contribution to the above integral in the 
limit $a\rightarrow 0$ is thus given by 
\begin{eqnarray}
&&2g^{2}[\int_{-\delta}^{\delta}\frac{d^{4}k}{(2\pi)^{4}} 
\frac{1}{H^{2}(k+ap)+(am\Gamma_{5})^{2}(k+ap)}
\frac{(aM)^{2}}{H^{2}(k+ap)+(aM)^{2}}\nonumber\\
&&\times\frac{1}{H^{2}(k)+(am\Gamma_{5})^{2}(k)}
\frac{(aM)^{2}}{H^{2}(k)+(aM)^{2}}]
\end{eqnarray}
for arbitrarily small but finite $\delta$. We can thus write 
this integral as 
\begin{eqnarray}
&&2g^{2}[\int_{-\delta}^{\delta}\frac{d^{4}k}{(2\pi)^{4}} 
\frac{1}{(k+ap)^{2}+(am)^{2}}
\frac{(aM)^{2}}{(k+ap)^{2}+(aM)^{2}}\nonumber\\
&&\times\frac{1}{k^{2}+(am)^{2}}
\frac{(aM)^{2}}{k^{2}+(aM)^{2}}]\nonumber\\
&&=2g^{2}[\int_{-\delta/a}^{\delta/a}\frac{d^{4}k}{(2\pi)^{4}} 
\frac{1}{(k+p)^{2}+m^{2}}
\frac{M^{2}}{(k+p)^{2}+M^{2}}\nonumber\\
&&\times\frac{1}{k^{2}+m^{2}}
\frac{M^{2}}{k^{2}+M^{2}}]\nonumber\\
&&\rightarrow
2g^{2}[\int_{-\infty}^{\infty}\frac{d^{4}k}{(2\pi)^{4}} 
\frac{1}{(k+p)^{2}+m^{2}}
\frac{M^{2}}{(k+p)^{2}+M^{2}}\nonumber\\
&&\times\frac{1}{k^{2}+m^{2}}
\frac{M^{2}}{k^{2}+M^{2}}]
\end{eqnarray} 
in the limit $a\rightarrow 0$. Here we used 
$H^{2}(k)\simeq k^{2}$ for $|k|<\delta$ by recalling the 
rescaling $ak_{\mu}\rightarrow k_{\mu}$. See Appendix.
This last expression is identical to
the continuum result in the higher derivative regularization.

The crucial aspect of this analysis is that the momentum 
variables are constrained to the infrared region in the 
limit $a\rightarrow 0$. Namely, the typical momentum variable
is constrained to be 
\begin{equation}
|ak^{\mu}| < \delta
\end{equation}
for arbitrarily small but finite $\delta$, which is a necessary 
condition for the validity of the Leibniz rule on the lattice.
\\ 

The  correction to the kinetic term of the scalar particle $\phi$
by the fermion loop diagram is given by
\begin{eqnarray}
g^{2}a^{2}Tr\phi^{\dagger}
\frac{H\Gamma_{5}}{H^{2}+(ma\Gamma_{5})^{2}}
\frac{(aM)^{2}}{H^{2}+(aM)^{2}}
\phi\frac{\Gamma_{5}H}{H^{2}+(ma\Gamma_{5})^{2}}
\frac{(aM)^{2}}{H^{2}+(aM)^{2}}
\end{eqnarray}
where the symbol $Tr$ includes the integral over the loop 
momentum as well as the trace over Dirac matrices.
The quadratic divergence and the logarithmic divergence 
associated with the mass term in this expression, when evaluated 
at vanishing external momentum, are cancelled by the scalar 
loop diagram (3.14) given below even for a finite $a$. 

By analyzing the infrared structure in the limit 
$a\rightarrow 0$, one can confirm  that the expression 
(3.12) is reduced to 
the continuum expression in the higher derivative regularization
\begin{equation}
g^{2}tr\int\frac{d^{4}k}{(2\pi)^{4}}\phi^{\dagger}(p)
\frac{\pslash+\kslash}{(k+p)^{2}+ m^{2}}
\frac{M^{2}}{(k+p)^{2}+M^{2}}
\phi(p)\frac{\kslash}{k^{2}+m^{2}}
\frac{M^{2}}{k^{2}+M^{2}}.
\end{equation}

The one-loop self-energy of the scalar particle $\phi$ produced 
by the scalar particle loop is given by 
\begin{eqnarray}
&&\frac{g^{2}}{2!}[F\phi^{2}
+(F\phi^{2})^{\dagger}]
[F\phi^{2}+(F\phi^{2})^{\dagger}]
\nonumber\\
&&\rightarrow
\frac{g^{2}}{2!}[8\phi\phi^{\dagger}\langle FF^{\dagger}\rangle
\langle\phi\phi^{\dagger}\rangle]\nonumber\\
&&=\frac{g^{2}}{2!}\int[8\phi\phi^{\dagger}
\frac{-H^{2}\Gamma_{5}^{2}}{H^{2}+(am\Gamma_{5})^{2}}
\frac{(aM)^{2}}{H^{2}+(aM)^{2}}
\frac{a^{2}}{H^{2}+(am\Gamma_{5})^{2}}
\frac{(aM)^{2}}{H^{2}+(aM)^{2}}]\nonumber\\
&&=-4g^{2}a^{2}\phi\phi^{\dagger}\int\Gamma_{5}^{2}
\frac{(aM)^{2}}{H^{2}+(aM)^{2}}
\frac{1}{H^{2}+(am\Gamma_{5})^{2}}\frac{(aM)^{2}}{H^{2}+(aM)^{2}}
\\
&&+\frac{(mg)^{2}a^{4}}{2!}\int[8\phi\phi^{\dagger}
\frac{\Gamma_{5}^{4}}{H^{2}+(am\Gamma_{5})^{2}}
\frac{(aM)^{2}}{H^{2}+(aM)^{2}}
\frac{1}{H^{2}+(am\Gamma_{5})^{2}}\frac{(aM)^{2}}{H^{2}+(aM)^{2}}].
\nonumber
\end{eqnarray}
The first term is quadratically divergent, and the remaining 
term is logarithmically divergent.
These terms, when evaluated at vanishing external momentum, 
precisely cancel the corresponding fermion contributions (3.12) 
even for a finite $a$.

The fermion self-energy correction is given by
\begin{eqnarray}
&&
4g^{2}a^{3}\bar{\psi}P_{+}[\int \frac{\Gamma_{5}H}
{H^{2}+(am\Gamma_{5})^{2}}\frac{(aM)^{2}}{H^{2}+(aM)^{2}}
\frac{1}{H^{2}+(am\Gamma_{5})^{2}}
\frac{(aM)^{2}}{H^{2}+(aM)^{2}}]P_{-}\psi\nonumber\\
&&+4g^{2}a^{3}\bar{\psi}P_{-}[\int \frac{\Gamma_{5}H}
{H^{2}+(am\Gamma_{5})^{2}}\frac{(aM)^{2}}{H^{2}+(aM)^{2}}
\frac{1}{H^{2}+(am\Gamma_{5})^{2}}
\frac{(aM)^{2}}{H^{2}+(aM)^{2}}]P_{+}\psi\nonumber\\
\end{eqnarray}
where we used the relation
\begin{equation}
P_{\pm}\Gamma_{5}H=\Gamma_{5}HP_{\mp}.
\end{equation}
By analyzing the infrared structure in the limit 
$a\rightarrow 0$, we again have 
\begin{eqnarray}
&&4g^{2}\bar{\psi}P_{+}[\int\frac{d^{4}k}{(2\pi)^{4}} 
\frac{\kslash+\pslash}
{(k+p)^{2}+m^{2}}\frac{M^{2}}{(k+p)^{2}+M^{2}}
\frac{1}{k^{2}+m^{2}}
\frac{M^{2}}{k^{2}+M^{2}}]P_{-}\psi\nonumber\\
&&+4g^{2}\bar{\psi}P_{-}[\int\frac{d^{4}k}{(2\pi)^{4}} 
\frac{\kslash+\pslash}
{(k+p)^{2}+m^{2}}\frac{M^{2}}{(k+p)^{2}+M^{2}}
\frac{1}{k^{2}+m^{2}}
\frac{M^{2}}{k^{2}+M^{2}}]P_{+}\psi.\nonumber\\
\end{eqnarray}
We have to examine if a universal wave function renormalization 
is sufficient to remove the divergences from these 3 
contributions (3.10), (3.13) and (3.17) in the limit 
$a\rightarrow 0$.
One can in fact show that a uniform subtraction of logarithmic 
infinity ( for $M\rightarrow$ large) renders all these 
expressions finite when renormalized at vanishing momentum. 
For the fermion contribution to the scalar kinetic term (3.13),
 we first rewrite it as
\begin{eqnarray}
&&4g^{2}\int\frac{d^{4}k}{(2\pi)^{4}}\phi^{\dagger}(p)\phi(p)
3!\int d\alpha d\beta d\gamma d\delta 
\delta(1-\alpha-\beta-\gamma-\delta)\times\\ 
&&\frac{-p^{2}(1-\alpha-\beta)M^{4}}{[p^{2}(\alpha+\beta)(1
-\alpha-\beta)+
\alpha(k^{2}+m^{2})
+\beta(k^{2}+M^{2})+\gamma(k^{2}+m^{2})+\delta(k^{2}+M^{2})]^{4}}
\nonumber
\end{eqnarray}
and similarly for the fermion self-energy correction (3.17). 
We then renormalize all the kinetic terms at $p=0$ by using the 
relation
\begin{eqnarray}
&&3!\int d\alpha d\beta d\gamma d\delta 
\delta(1-\alpha-\beta-\gamma-\delta)\nonumber\\ 
&&\times\frac{1-\alpha-\beta}{[\alpha(k^{2}+m^{2})
+\beta(k^{2}+M^{2})+\gamma(k^{2}+m^{2})+\delta(k^{2}+M^{2})]^{4}}
\nonumber\\
&&=3!\int d\alpha d\beta d\gamma d\delta 
\delta(1-\alpha-\beta-\gamma-\delta)\nonumber\\ 
&&\times\frac{1/2}{[\alpha(k^{2}+m^{2})
+\beta(k^{2}+M^{2})+\gamma(k^{2}+m^{2})
+\delta(k^{2}+M^{2})]^{4}}.
\end{eqnarray}
In the one-loop level, we can thus maintain supersymmetry 
including renormlization factors\footnote{Here we are repeating 
the known analysis in continuum theory\cite{iliopoulos}.} in the 
limit $a\rightarrow 0$.
In other words, all the supersymmetry breaking terms for  
finite lattice spacing should vanish for $a\rightarrow 0$. 

\subsection{Two and higher-loop diagrams}

\subsubsection{Tadpole diagrams}
We start with the analysis of tadpole diagrams. One can confirm 
that the tadpole diagrams for the auxiliary field $F$ in the 
two-loop level precisely
cancel even for a finite lattice spacing $a$, and similarly for
the field $F^{\dagger}$.\\

For the scalar field $\phi$, we have 4 tadpole diagrams in the 
two-loop level. These diagrams do not quite cancel for a 
finite $a$ due to
the failure of the Leibniz rule for general momenta.
But for non-vanishing contributions in the limit 
$a\rightarrow 0$, one can reduce the Feynman amplitudes to 
those of the continuum theory 
with higher derivative regularization, which are then shown to 
cancel precisely. We can thus maintain the vanishing tadpole
diagrams in the two-loop level.

For example, we have a two-loop tadpole contribution arising 
from a fermion loop diagram with a scalar exchange correction, 
which contains a quadratic divergence in a naive sense,
\begin{eqnarray}
&&4\phi g^{3}\int_{-\pi}^{\pi}\frac{d^{4}p}{(2\pi)^{4}}
\int_{-\pi}^{\pi}\frac{d^{4}k}{(2\pi)^{4}} 
tr\{\frac{a\Gamma_{5}H(p)}{H^{2}(p)+(am\Gamma_{5}(p))^{2}}
\frac{M^{2}}{H^{2}(p)+(aM)^{2}}\nonumber\\
&&\times
\frac{a^{2}m(\Gamma_{5}(p))^{2}}{H^{2}(p)+(am\Gamma_{5}(p))^{2}}
\frac{M^{2}}{H^{2}(p)+(aM)^{2}}\nonumber\\
&&\times
\frac{a\Gamma_{5}H(k+p)}{H^{2}(k+p)+(am\Gamma_{5}(k+p))^{2}}
\frac{M^{2}}{H^{2}(k+p)+(aM)^{2}}\nonumber\\
&&\times\frac{a^{2}}{H^{2}(k)+(am\Gamma_{5}(k))^{2}}
\frac{M^{2}}{H^{2}(k)+(aM)^{2}} \} 
\end{eqnarray}
where we used the rescaled variables 
$ap_{\mu}\rightarrow p_{\mu}$ and $ak_{\mu}\rightarrow k_{\mu}$.
 By analyzing the limit $a\rightarrow 0$,
one can confirm that only the infrared regions 
\begin{equation}
|p_{\mu}|<\delta, \ \ \ \ \ |k_{\mu}|<\delta \ \ \ \ for\ \ \  
all \ \ \ \mu
\end{equation}
with arbitrarily small but finite $\delta$ give a non-vanishing
contribution
\begin{eqnarray}
&&4\phi g^{3}\int_{-\delta}^{\delta}\frac{d^{4}p}{(2\pi)^{4}}
\int_{-\delta}^{\delta}\frac{d^{4}k}{(2\pi)^{4}} 
tr\{\frac{a\Gamma_{5}H(p)}{H^{2}(p)+(am\Gamma_{5}(p))^{2}}
\frac{M^{2}}{H^{2}(p)+(aM)^{2}}\nonumber\\
&&\times
\frac{a^{2}m(\Gamma_{5}(p))^{2}}{H^{2}(p)+(am\Gamma_{5}(p))^{2}}
\frac{M^{2}}{H^{2}(p)+(aM)^{2}}\nonumber\\
&&\times
\frac{a\Gamma_{5}H(k+p)}{H^{2}(k+p)+(am\Gamma_{5}(k+p))^{2}}
\frac{M^{2}}{H^{2}(k+p)+(aM)^{2}}\nonumber\\
&&\times\frac{a^{2}}{H^{2}(k)+(am\Gamma_{5}(k))^{2}}
\frac{M^{2}}{H^{2}(k)+(aM)^{2}} \}. 
\end{eqnarray}
By rescaling the momentum variables to original ones and 
considering the limit $a\rightarrow 0$, one recovers the 
continuum result in the higher derivative regularization.
Other tadpole amplitudes in the two-loop level are  analyzed 
similarly.

\subsubsection{General diagrams in two and higher-loop level}

Similarly, one can confirm that the non-vanishing parts of all 
the Feynman amplitudes in the two-loop level in the limit 
$a\rightarrow0$   are reduced to those 
of the continuum theory with higher derivative regularization.
Namely, all the loop momenta in the Feynman amplitudes are 
constrained to be in the infrared region, where the Leibniz 
rule is satisfied. 

To the extent that the non-renormalization and 
other good properties are maintained in the continuum theory 
with higher derivative regularization\cite{iliopoulos}, our 
lattice regularization thus reproduces all the good properties 
of the supersymmetric Wess-Zumino model in the limit 
$a\rightarrow 0$.

One can extend  this analysis up to any finite order in 
perturbation theory, since the power counting in this 
regularized theory is effectively superconvergent; the higher
order diagrams are thus more ultra-violet convergent and thus 
less sensitive to the lattice cut-off for $a\rightarrow 0$.
All the supersymmetry breaking terms induced by the failure 
of the Leibniz rule are thus shown to vanish in the limit 
$a\rightarrow 0$.

\subsection{Non-perturbative treatment}
As for the non-perturbative treatment of our model, one may first
perform the path integral over the Majorana fermion which 
produces the Pfaffian 
\begin{eqnarray}
&&\int{\cal D}\phi{\cal D}\phi^{\dagger}{\cal D}F
{\cal D}F^{\dagger}\times\nonumber\\
&&\sqrt{\det[\frac{1}{\Gamma_{5}}
\frac{H}{a}\frac{(H^{2}+(aM)^{2})}{(aM)^{2}} 
+ m\frac{(H^{2}+(aM)^{2})}{(aM)^{2}}
+2g(P_{+}\phi P_{+}+P_{-}\phi^{\dagger}P_{-})]}
\nonumber\\
&&\times\exp\{\int[-\phi^{\dagger}\frac{H^{2}}{a^{2}}
\frac{(H^{2}+(aM)^{2})}{(aM)^{2}}\phi
+F^{\dagger}\frac{1}{\Gamma^{2}_{5}}
\frac{(H^{2}+(aM)^{2})}{(aM)^{2}}F\nonumber\\
&&
+m[F\frac{(H^{2}+(aM)^{2})}{(aM)^{2}}\phi
+(F\frac{(H^{2}+(aM)^{2})}{(aM)^{2}}\phi)^{\dagger}]
+g[F\phi^{2}+(F\phi^{2})^{\dagger}]]\}\nonumber\\  
&&=\int{\cal D}\phi{\cal D}\phi^{\dagger}{\cal D}F
{\cal D}F^{\dagger}\times\nonumber\\
&&\sqrt{\det[
-\frac{H}{a}\frac{(H^{2}+(aM)^{2})}{(aM)^{2}} 
+ m\frac{(H^{2}+(aM)^{2})}{(aM)^{2}}\Gamma_{5}
+2g(P_{+}\phi P_{+}+P_{-}\phi^{\dagger}P_{-})\Gamma_{5}]}
\nonumber\\
&&\times\exp\{\int[-\phi^{\dagger}\frac{H^{2}}{a^{2}}
\frac{(H^{2}+(aM)^{2})}{(aM)^{2}}\phi
+F^{\dagger}\frac{(H^{2}+(aM)^{2})}{(aM)^{2}}F\\
&&
+m[F\Gamma\frac{(H^{2}+(aM)^{2})}{(aM)^{2}}\phi
+(F\Gamma\frac{(H^{2}+(aM)^{2})}{(aM)^{2}}\phi)^{\dagger}]
+g[F\Gamma\phi^{2}+(F\Gamma\phi^{2})^{\dagger}]]\}\nonumber
\end{eqnarray}
where in the second expression we rescaled the field variables 
as 
\begin{equation}
F\rightarrow \Gamma F, \ \ \ \ F^{\dagger}\rightarrow 
F^{\dagger}\Gamma
\end{equation}
with $\Gamma\equiv \sqrt{\Gamma^{2}_{5}}=\sqrt{1-H^{4k+2}}$. In 
the last 
expression in (3.23) we have no singularity associated with 
$1/\Gamma_{5}$.
This path integral (or after performing the path integral 
over $F$ and $F^{\dagger}$ ) may be evaluated non-perturbatively.
Since the Wess-Zumino model is not asymptotically free, the 
non-perturbative result in the continuum limit $a\rightarrow 0$
may be defined with a finite $M$, which provides a finite
mass scale to specify the renormalized parameters; one can thus
prevent the coupling constant from increasing indefinitely. In 
this sense our possible non-perturbative formulation, which is 
inferred from perturbative considerations, is consistent. 

If the above path integral (3.23), when evaluated 
non-perturbatively, gives a well-defined result in the limit
$a\rightarrow 0$, it is expected that the path integral 
 defines a theory which incorporates
the quantum effects up to the energy scale $M$. The path 
integral may then be effective in defining a
non-perturbative low energy effective theory\footnote{If one 
should be able to evaluate the continuum theory with higher 
derivative regularization in a non-perturbative way, one would 
obtain the same result. } for 
\begin{equation} 
|p_{\mu}|\sim m \ll M
\end{equation}
where $p_{\mu}$ is the typical external momentum carried by 
field variables.

\section{Discussion}

We have discussed a way to ensure the Leibniz rule for 
the supersymmetric Wess-Zumino model on the lattice. The 
basic observation is that the lattice Leibniz rule is reduced 
to that of continuum theory if the generic  momentum $k_{\mu}$
carried by any field variable is constrained in the infrared 
region $|ak_{\mu}|<\delta$ for arbitrarily small but finite 
$\delta$ in the limit $a\rightarrow 0$. A way to ensure this 
momentum condition is to apply the higher derivative 
regularization to the Wess-Zumino model so that the theory 
becomes finite up to any finite order in perturbation theory.
On the basis of the analysis of Feynman amplitudes, we have 
shown that this is in fact realized in our lattice 
formulation  which incorporates a lattice version of higher 
derivative regularization. 
All the supersymmetry breaking terms induced by the failure of
the Leibniz rule thus become irrelevant 
in the sense that they all vanish in the limit $a\rightarrow 0$.
We suggested that this mechanism may work even in a 
non-perturbative sense.

In this analysis, it is crucial that the 
Ginsparg-Wilson operator is free of species doublers so that 
the lattice operator $H$ is of order $O(1)$ in the momentum 
regions of the would-be species doublers\footnote{It is possible
to implement the present mechanism for the model based on the 
Wilson fermion\cite{bartels}, if one defines the higher 
derivative regulator suitably by including the Wilson term.
The non-perturbative analysis would, however, become more 
involved since the basic symmetries such as chiral symmetry and 
holomorphicity are spoiled.}. 
Besides, the 
Ginsparg-Wilson operator maintains some of the basic symmetries 
such as chiral symmetry, $U(1)\times U_{R}(1)$ symmetry and 
holomorphicity. 

In conclusion, we have presented a possible way to maintain
the Leibniz rule on the lattice. It is yet to be seen if this 
analysis is extended to supersymmetric Yang-Mills 
theories on the lattice\cite{banks}-\cite{maru}, though we 
naively expect that supersymmetric theories which are 
perturbatively finite ( such as $N=4$ theory ) may be put on 
the lattice consistently.
One may also want to find a more drastic way to overcome the 
difficulty associated with the Leibniz rule, which might 
lead to a completely new understanding of lattice regularization.

\appendix
\section{Ginsparg-Wilson operators}

An explicit form of the general Ginsparg-Wilson 
operator $H$ \cite{fujikawa}, which satisfies 
the algebra (2.3), is given in momentum space by
\begin{eqnarray}
H(ap_{\mu})&=&\gamma_{5}(\frac{1}{2})^{\frac{k+1}{2k+1}}
(\frac{1}{\sqrt{H^{2}_{W}}})^{\frac{k+1}{2k+1}}
\{(\sqrt{H^{2}_{W}}+M_{k})^{\frac{k+1}{2k+1}}
-(\sqrt{H^{2}_{W}}-M_{k})^{\frac{k}{2k+1}}
i\frac{\sslash}{a} \}\nonumber\\
&=&\gamma_{5}(\frac{1}{2})^{\frac{k+1}{2k+1}}
(\frac{1}{\sqrt{F_{(k)}}})^{\frac{k+1}{2k+1}}
\{(\sqrt{F_{(k)}}+\tilde{M}_{k})^{\frac{k+1}{2k+1}}
-(\sqrt{F_{(k)}}-\tilde{M}_{k})^{\frac{k}{2k+1}}
i\sslash \}\nonumber\\
&&
\end{eqnarray}
where $k$ is a non-negative integer and 
\begin{eqnarray}
F_{(k)}&=&(s^{2})^{2k+1}+\tilde{M}_{k}^{2},\nonumber\\
\tilde{M}_{k}&=&[\sum_{\mu}(1-c_{\mu})]^{2k+1}
-m_{0}^{2k+1}
\end{eqnarray}
with
\begin{eqnarray}
&&s_{\mu}=\sin ap_{\mu}\nonumber\\
&&c_{\mu}=\cos ap_{\mu}\nonumber\\
&&\sslash=\gamma^{\mu}\sin ap_{\mu}.
\end{eqnarray}
This operator is known to be local 
and free of species doublers\cite{fujikawa3}, and 
 this operator for $k=0$ is reduced to  Neuberger's overlap 
operator\cite{neuberger}. 
Our Euclidean Dirac matrices are hermitian, 
$(\gamma^{\mu})^{\dagger}=\gamma^{\mu}$, and the inner 
product is defined to be $s^{2}\geq 0$. Note that $H^{2}$
( and consequently $\Gamma_{5}^{2}=1-H^{4k+2}$ ) is independent 
of Dirac matrices.
The parameter $m_{0}$ is constrained by $0<m_{0}<2$ to avoid 
species doublers, and 
$2m^{2k+1}_{0}=1$ gives a proper normalization
of $H$, namely, for an infinitesimal $p_{\mu}$, i.e.,
for $|ap_{\mu}|\ll 1$,
\begin{equation}
H\simeq-\gamma_{5}ai\pslash(1+O(ap)^{2})
+\gamma_{5}(\gamma_{5}ai\pslash)^{2k+2}
\end{equation}
to be consistent with $H=\gamma_{5}aD$; the last term in the 
right-hand side is the 
leading term of chiral symmetry breaking terms.

We thus have
\begin{equation}
H^{2}=1
\end{equation}
just on top of the would-be species doublers, for example,
$ap_{\mu}=(\pi,0,0,0)$, and 
\begin{equation}
H^{2}\simeq (ap_{\mu})^{2}
\end{equation}
for $|ap_{\mu}|\ll 1$ independently of the parameter $k$.

\end{document}